\documentstyle[epsf,graphicx]{mn}

\def\x2{$\chi^{2}$}

\def\ginga{{\it Ginga}}

\newbox\grsign \setbox\grsign=\hbox{$>$} \newdimen\grdimen \grdimen=\ht\grsign
\newbox\simlessbox \newbox\simgreatbox \newbox\simpropbox
\setbox\simgreatbox=\hbox{\raise.5ex\hbox{$>$}\llap
     {\lower.5ex\hbox{$\sim$}}}\ht1=\grdimen\dp1=0pt
\setbox\simlessbox=\hbox{\raise.5ex\hbox{$<$}\llap
     {\lower.5ex\hbox{$\sim$}}}\ht2=\grdimen\dp2=0pt
\setbox\simpropbox=\hbox{\raise.5ex\hbox{$\propto$}\llap
     {\lower.5ex\hbox{$\sim$}}}\ht2=\grdimen\dp2=0pt

\def\ginga{{\it Ginga~}}

\begin{document}

\title[Monitoring {\it RXTE} Observations of Markarian 348 ]
{Monitoring {\it RXTE} Observations of Markarian 348: the origin of the column density 
variations }

\author[Akylas et al.]
{\Large A. Akylas$^{1,2}$,  I. Georgantopoulos$^{1}$, 
R. G. Griffiths$^3$, I. E. Papadakis$^4$, A. Mastichiadis$^{2}$, \\
\Large R. S. Warwick$^3$,  K. Nandra$^5$, D.A. Smith$^6$\\    
$^1$ Institute of Astronomy \& Astrophysics, National Observatory of Athens, I. Metaxa
 B. Pavlou, Penteli, 15236, Athens, Greece \\
$^2$ Physics Department University of Athens, Panepistimiopolis, 
 Zografos, 15783, Athens, Greece \\  
$^3$ Department of Physics and Astronomy, University of Leicester, Leicester LE1 7RH \\
$^4$ Physics Department University of Crete, 73010, Heraklion, Greece \\
$^5$ Laboratory for High Energy Astrophysics, Code 660, NASA/Goddard Space Flight 
Center, Greenbelt, MD20771, U.S.A. \\
$^6$ University of Maryland, College Park, MD 20742,U.S.A.
}

\maketitle

\begin{abstract}

We analyze 37 {\it RXTE} observations of the type 2 Seyfert galaxy Mrk348 
obtained during a period of 14 months. 
We confirm the spectral variability previous reported by Smith et al., 
in the sense that the
column density decreases by a factor of $\sim$ 3 as the count rate 
increases.  
 Column density variations could possibly originate  either
due to the random drift of clouds within the absorption screen, 
or due to photoionization processes.
Our  modeling of  the observed variations implies that the first 
scenario is more likely.   
 These clouds should lie in a distance of $>$2 light years from the 
source, having a diameter of a few  light days and a density of 
$>10^7~ \rm cm^{-3}$, hence probably residing outside the 
 Broad Line Region. 
 
\end{abstract}


\newpage

\section{Introduction}

 Monitoring observations are a powerful  
 tool in the study  of the nuclear environment of type 2  Seyfert galaxies.
 Although the presence of X-ray flux variability in Seyfert-2 galaxies 
 is well established   (eg Georgantopoulos \& Papadakis 2001 and references 
therein),
  our understanding of their spectral variations remains limited. 
 Nevertheless, monitoring of the X-ray spectra of 
 a few bright Seyfert-2 galaxies 
 have revealed some intriguing results.
  For example,  Warwick et al. (1988) first reported  
 variations of the column density in ESO 103-G35. Using {\it EXOSAT} data, 
they found a decrease in the column density by a factor of $\sim1.7$ in a period of 
90 days. Warwick et al. (1993) found column density variations in NGC 7582 
 by a factor of $\sim$ 3 over an interval of about 4 years, using \ginga data.
 They attributed these variations to motions of clouds near the 
 central source.    
 Investigation of the spectrum of NGC 7582 based on
{\it ASCA} data by Xue et al. (1998) confirmed the existence
of significant column density  variations (by a factor of $\sim 2$) 
 over a timescale of 2  years. 
 Variability analysis of column density variations in large samples of Seyfert-2 
 galaxies has been performed by Malizia et al.  (1997) and Risaliti, Elvis \& Nicastro 
 (2002) using mainly literature data. 
 They find  variation of the column density in time scales
 of a few months up to several years.   

  Systematic monitoring observations only became feasible with the {\it RXTE } 
 mission. In particular, 
 Georgantopoulos et al. (1999), Georgantopoulos \& Papadakis (2001),
 Smith, Georgantopoulos \& Warwick (2001) present monitoring observations 
 of several Seyfert-2 galaxies (Mrk3, ESO 103-G35, IC 5063, NGC 4507, 
 NGC 7172 and Mrk 348) spanning time periods from about seven days 
 to seven months. They found statistically significant spectral variations
in all cases. In some objects the variations appear to be  caused 
by intrinsic power-law 
slope changes whereas in others column density variations dominate.

 In this paper we present an analysis of 37 {\it RXTE} observations of 
 the Seyfert 2 galaxy Mrk348.  
Previously Smith et al. (2001) have reported the results
 from 12 $RXTE$  observations covering a period of six 
 months. Here we use an expanded sample of 37 $RXTE$  
 monitoring observations of Mrk 348 (including the  
12 reported by Smith et al. 2001), spanning a  time interval of 
14 months, to investigate further the nature of  the X-ray
spectral variability exhibited by this source.  
 More specifically, we investigate whether 
photoionization of the absorbing screen 
or alternatively motion of clouds along the line of sight can 
 reproduce the observed spectral variability.

\section{The Data}

We use 37 observations of  Mrk348  
obtained with the Rossi X-ray Timing Explorer $(RXTE)$ 
mission. The data are spanning a period of $\sim$ 14 months 
from May 24, 1996 to July 12, 1997. 
Each observation lasts typically ~2500-5000 sec. Here we present data 
from the Proportional Counter Array (PCA, Glasser, Odell \& Seufert 1994)  
instrument only. We use PCA units 0 to 2. The data from the other two units 
were discarded as these  were turned off on some occasions. We extract 
3-20 keV spectra from only the top layer in order to maximize the 
signal to noise ratio. The data were selected using the standard screening
criteria. The background subtraction was done using the PCABACKEST 
v2 routine of FTOOLS which takes into account both cosmic and internal
background using the latest L7 model (see Smith et al. 2001
for a detailed description).  

The background subtracted light curve  is presented in Fig 1. 
In particular the count rate of the source increases by a factor of 
$\sim5$ in the last few months. In the 1996 data the 
variations present half the above  amplitude
in a period of $\sim15$ days.

\begin{figure}
\rotatebox{0}{\epsfxsize=9cm \epsffile{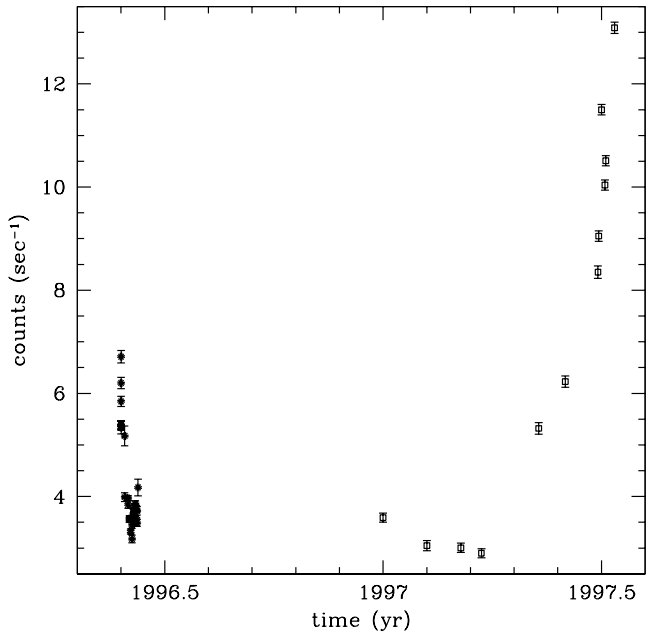}}      
\caption{The background subtracted count rate versus time 
for the 37 observations of Mrk348. The open squares denote the observations of
Smith et al. (2001)}  
\end{figure}

\section{Spectral Analysis} 

\subsection{Neutral absorption models}
We use the XSPEC v11.0 software package in our spectral fitting analysis. 
In order to improve the signal-to-noise ratio, 
we combine individual spectra obtained within up to four days. 
In this way we obtain 13 spectral datasets (9 of which overlap with the data
analysed by Smith et al. 2001). 
Errors quoted correspond to the 90 per cent confidence level for 
one interesting parameter. All the energies refer to the observer's rest frame.

We perform  joint fits in the 13 combined observations in order to derive the 
best fit model. We  fit the data using a simple absorbed power-law model plus a 
Gaussian line ($\sigma=0.1$ keV) to account for Fe K emission. 
Both the column density and the photon 
index  were tied to a single value  and only the normalization of 
the power-law was 
allowed to  vary freely. This model provides an unacceptable fit 
to the data ($\chi^2=2310/568$). Therefore we  set  the $N_H$ parameter free 
for all the observations. We obtain a  good fit with $\chi^2=480/556$. Furthermore 
we investigate possible variations of the photon index. We keep $\rm N_H$ 
tied to a single value and let the photon index parameter to vary free 
for all the  observations. 
The value of $\chi^2$ (538/556) implies that this is a good fit.
We also allow both the column density and the photon index parameters  
to vary freely. 
This gives a statistically significant reduction in $\chi^2$ (390/544).   
We conclude that both column density and photon index variations  exist.

Apparent  variations of the photon index  could be due to the presence of a  
reflection component (see for example Georgantopoulos et al. 1999 in 
the case of  Mrk3). 
To test this assumption  we include in our model a reflection component 
({\sl PEXRAV} model in XSPEC, R=-1, $\rm E_{cut}$=150 keV, solar abundances). 
We repeat the same analysis as above. First we 
let only the column density and the power-law normalization  to 
vary freely. The power-law and the reflection component photon index 
were tied to a single value for all the observations. The normalization of the 
reflection component was tied to a single value.  
We obtain a good fit with $\chi^2=365/555.$ 
The best fit photon index value  was $1.85^{+0.02}_{-0.02}$.
 Therefore, in the subsequent analysis,
  we fix the reflection photon index to 1.85. In addition to $N_H$
we also let the power-law photon index to vary freely.
We find $\chi^2=350/543$.  
The reduction in $\chi^2$ was less than 15 for 12 additional parameters compared 
to the previous model ($\Gamma$ fixed to a common value and $N_H$ free) 
which is not statistically significant.
We also let the normalization of either the Fe line or 
 the reflection component to vary freely between the 
 observations. The fit is  not improved at a statistically significant level:
 $\Delta \chi^2\approx$ 6 and 10 respectively for 12 additional parameters. 
 Therefore we conclude that the only varying parameters between the 
observations were the  power-law normalization and the column density. 

Hereafter, we call the variable column density absorbed power-law plus a 
Gaussian line and a constant reflection component as the "standard" model.
In Fig. 2 we plot the column density variations as a function of the 
observation time. 
The best fitting model parameters are listed in Table 1. They are in good 
agreement with the results of Smith et al. (2001).
The ratio of the reflected to the total flux
( 3 to 20 keV) varies between 0.09 and 0.3. 
 The best fitting line energy is 
 E=$6.1^{+0.1}_{-0.1}$ keV ($\sigma$=0.1 keV).
 This energy is below that expected 
 for cold iron. As noted by Smith et al. (2001) the discrepancy could be caused by  
 uncertainties in the instrument response around the XeL edge.
The EW of the Fe line varies between 70 and 340 eV, being anti-correlated 
 with the power-law flux variations.

 Recently, evidence for a double absorbing screen   which partially
  or fully covers the  
 nucleus in several type 2 Seyfert galaxies has emerged
 (eg Turner et al. 2000). 
 Smith et al. (2001) found that a double screen model consisting of a 
variable column density and a fixed partial coverer 
 provides a good description to the data of Mrk348.       
Here,  we  use a slightly different screen absorption 
model consisting of a neutral variable covering fraction and a constant 
column density (models {\sl ZPCFA} and {\sl WA} in XSPEC). 
  In this model, the first  
 variable screen could account for   clouds orbiting  close to the
 nucleus while the outer constant  screen may be
 associated with the molecular  torus.
The best fitting results for this  model are listed in Table 1.   
 The $\chi^2$ value (322/552 d.o.f)  implies that this model provides the best
 description to our data so far.  
  The constant column density  is  
 $\rm N_H=12^{+1}_{-1} \times10^{22}~ cm^{-2}$ while  the variable 
 screen has $\rm  N_H=28^{+1}_{-1} \times10^{22}~ cm^{-2}$ and 
 a covering fraction between 0.05 and 1.

\subsection{Photoionization models}

Fig. 3 shows the best fitting column density values
for the standard model as a function of the source unobscured 
 luminosity (crosses).
There appears to be an anti-correlation 
 in the sense that the column density 
decreases with increasing flux.
 This  anti-correlation is introduced mainly by the last three 
 points and is found to be significant at the 90 percent confidence level 
 using the Kendall's test (Press et al. 1986). 
Such a behaviour suggests that the column density variations  
 may be caused by photoionization processes.
 In order to check this possibility we use the 
ionized absorber model {\sl ABSORI}
(Magdziarz \&  Zdsiarski 1985) in XSPEC instead of the 
neutral  partial covering fraction absorption model {\sl ZPCFA}. 
 
 We use again a double absorption screen model. Both  absorbers are assumed 
 to fully cover  the source. One absorber is neutral  
(with column  density $\rm N_H^1$) while the other  ($\rm N_H^2$)   
 is partially  ionized and variable as   $\rm \xi=L/nR^{2}$. Hence the 
{\it effective} column density ($\rm N_H^2$) varies with luminosity. We also
assume $\Gamma=1.85$, a  Gaussian line (E=6.1 keV and $\sigma=0.1$ keV) and 
an edge at 5.38 keV.

The best fitting results are listed in Table 1.    
The  neutral and ionized  column densities are  
$\rm N_H^1\sim 8 \times 10^{22}~  cm^{-2}$ and
$\rm N_H^2\sim 25 \times 10^{22}~  cm^{-2}$ respectively. 
The ionization parameter varies 
roughly between 0 and 450 while the best fitting  temperature 
is  $\sim 5 \times 10^5$ K.  Although this model provides 
an acceptable fit to the data ($\chi^2$=450 for 553 d.o.f), it yields 
 an increase in $\chi^2$ by $\sim 130$ compared to 
the double neutral absorption model discussed earlier.
 
We have cross-checked the above results using the photoionization code XSTAR
(Kallman \& Krolik 2000) instead of the $\sl ABSORI$ model. The XSTAR  
calculates the physical conditions, emission 
and absorption spectra of photoionized gases.   
We created  absorption  table models representing 
 the fraction of incident continuum 
emerging  from a  photoionized screen with a range of photoionization
parameters,  (log$\xi$=-3 - 3) and column densities 
(\rm $N_H = 6 \times 10^{22}-4 \times 10^{23}~ \rm cm^{-2}$). A constant 
density of n=$10^9$
hydrogen atoms/ions $\rm cm^{-3}$ was used and  the gas temperature was 
fixed at $10^6$ K. 
 We fit the data using the same model as above, 
substituting the $\sl ABSORI$ 
component  with the absorption table model produced by XSTAR. 
 The best fitting  values  are listed in Table 1.
The value of the   neutral column density is 
$\rm N_H^1\sim 8 \times 10^{22}~  cm^{-2}$ and the value of the total 
  column density subject to photoionization is  $\rm N_H^2\sim 27 \times
10^{22}~ cm^{-2}$. The ionization parameter varies roughly between
0.001 and 350. These values are in agreement with the $\sl ABSORI$ 
 model results.

Both {\sl ABSORI} and XSTAR  results suggest that in the  photoionization 
scenario the medium subject to the  intense photoionizing flux
has a  column density of   $\sim 27 \times 10^{22} \rm cm^{-2}$.  
In  Fig. 3 the solid line represents the column density of a neutral 
gas giving the equivalent absorption (within the limitations of both
the {\it RXTE} spatial resolution and the crude modelling) 
as the  photoionized medium.
In order  to convert the ionization parameters  ($\rm \xi=L/nR^2$) 
derived  from XSTAR to luminosities, 
we have assumed $\rm nR^2 \approx 10^{42}~ cm^{-1}$.
Different values of $\rm nR^2$ will move the theoretical curve in Fig. 3 
along the x-axis.
 
Although the relatively poor $\chi^2$ 
values we obtain  rather argue against 
 photoionization models, the key test is the presence of 
 absorption edges due to ionized Fe. In these models, the observations with 
low effective  $\rm N_H$ should correspond 
to a highly ionized state. 
Therefore ionized edges should be present in their spectra. 
We fit (using the "standard" model)
 all the observations separately in order to 
investigate the existence of an ionized   edge. The energy of the edge 
was allowed to vary between 7.1 and 8 keV.
Even in the case where the column density was at the lowest state
we find no evidence for such a component 
($\Delta \chi^2<1,~ \rm E_{edge}=7.26 keV$ and $0<\tau<0.1)$.   
However, this could be due  to the relatively low resolution  
of the PCA which makes difficult to test the presence of 
such features.  
Therefore we checked for an Fe edge in the {\it ASCA}  observation of  Mrk348, 
obtained in  August 4, 1995. We fit simultaneously both 
SIS and GIS spectral files (exposure time $\sim$ 50 ks)  obtained from the  
Tartarus AGN database  (Turner et al. 1998).
The best fit model  ($\rm \chi^2=220/180$ d.o.f) 
 consists of a  power-law with
 $\rm \Gamma=0.5^{+0.2}_{-0.2}$ absorbed by a column density of
 $\rm N_H \sim 7\times 10^{22}~ cm ^{-2}$,
plus a Gaussian line at 6.4 keV ($\rm \sigma=0.01~ keV$). 
There is  no statistically significant evidence for the existence of 
an ionized  edge. 
 Indeed, we obtain $\Delta \chi^2<3$ for two additional parameters when
 such a feature is included in the fit. The best fitting values are 
$\rm E=7.2^{+1.0}_{-1.7}~keV$ and $\tau=0.20^{+0.15}_{-0.15}$.

\begin{figure}
\rotatebox{0}{\epsfxsize=9cm \epsffile{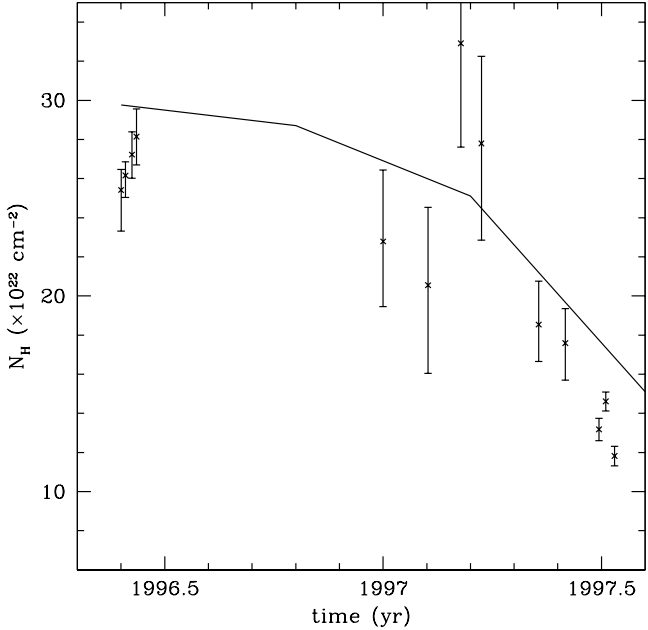}}   
\caption{The $\rm N_H$ variations of the standard model (crosses) 
plotted as a function of the observation time. The solid line model describes  
the expected  column density variations due to the circular motion 
of a spherical cloud around the source }
\end{figure}

\begin{figure}
\rotatebox{0}{\epsfxsize=9cm \epsffile{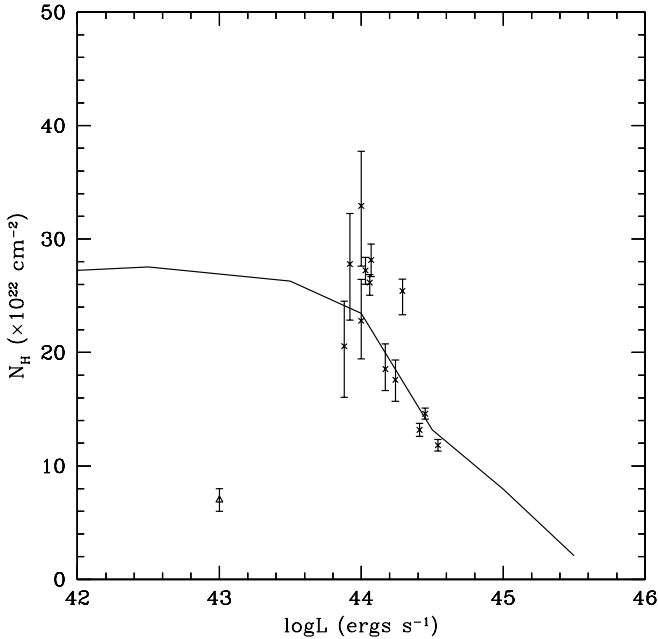}}   
\caption{The observed column density variations, obtained from the 
standard model as a function of the unobscured luminosity in the 0.0136-13.6 keV 
band (crosses). The solid line denotes
the XSTAR theoretical curve  describing the changes in the 
column density due to ionization.  
The triangle shows the results from the 1995 {\it ASCA}  observation using the 
standard model }

\end{figure}

\subsection{Cloud models}

Assuming, based on the above results, that a photoionization scenario 
alone cannot 
properly describe the observed column density variations we further 
try to derive  
constraints on the size, the velocity 
and the density of the neutral clouds which 
could alternatively be responsible for the $\rm N_H$ variations.
More specifically, we assume that the column density 
 variations are caused by the motion of a {\it single} cloud through 
 the line of sight. We further assume that this is a spherical, (with diameter D)  
 neutral or weakly ionized  cloud, ($\xi \sim 1$) moving 
 in a keplerian circular orbit  
 around the central mass. Then, the velocity and crossing time of the 
 cloud are given by $v=(GM/r)^{1/2}$ and $t=D/2v$. 
 Combining the above two equations with the definition of the 
 ionization parameter,    
 the density and the radius of this cloud are given by 
$n_{10}=129 \Delta N_{22}^{4/5} L_{42}^{1/5}/(t^{4/5} M_{8}^{2/5} \xi^{1/5})$ 
 and  
$r_{16}=0.09 t^{2/5} M_{8}^{1/5} L_{42}^{2/5}/(\Delta N_{22}^{2/5} \xi^{2/5})$, 
 (Griffiths 1999, Risaliti et al. 2002) where
 $n_{10}$ is the density of the cloud in units of $10^{10} ~\rm cm^{-3}$,
$r$ is the radius of its orbit in units of $10^{16} ~\rm cm$, 
 $\Delta N_{22}$ 
is the change in the column density in units of $10^{22} ~\rm cm^{-2}$
, $t$ is the time scale of the $\rm N_H$ variations, L is the 
unobscured luminosity in units of  $10^{42}~\rm ergs~s^{-1}$ and    
 $M_8$ is the mass of the  black hole in units of $10^8$ $M_\odot$.
 The mass of Mrk348 (Nishiura \& Taniguchi 1998)
 is  $5\times 10^8$ $M_\odot$. 
 For t=$\rm 4\times10^{7}~sec$,  
$\rm \Delta N_{22}=14$ (based on Fig. 2) and L=$50$,    
 we obtain $n\sim 10^7~ \rm cm ^{-3}$ and $r\sim 2.3$ light years.  
 Then $v\sim 1700~\rm km~s^{-1}$ and $D_{cloud}\sim 0.014$ light years. 
In Fig. 2 the solid line describes  the expected column density variations 
due to the motion of a spherical cloud with the above properties 
 in the line of sight. 
 In this model a constant column density of $\rm \sim 10^{23}~ cm^{-2}$,
(due to the constant absorption screen found in 
 the best fit model) was also added.  
 Note that the above calculations 
 give only a rough estimate for the   distance of 
 the cloud.  In reality, the value of the ionization parameter remains unknown 
 and introduces some uncertainty in the estimate of the 
 cloud distance. Using a lower limit value for $\xi$ of $10^{-3}$
 we find upper limits of $r\sim37$ light years  
and  $n\sim 4 \times 10^7~ \rm cm ^{-3}$ for the distance 
 and the density respectively. 
 The Broad Line Region (BLR) in Active Galactic Nuclei lies 
 approximately from several light days to several light months and its density 
 is $\rm \sim 10^9 cm^{-3}$. 
The velocities in the BLR vary between 1000-10000 km s$^{-1}$.
The Narrow Line Region lies in much larger distances (up to 1 kpc) and its 
density is  $\rm \sim 10^3 cm^{-3}$. Characteristic velocities of this region
are $\rm \sim 500~km~s^{-1}$ (Peterson 1993). 
 Therefore in the case of Mrk348 the clouds responsible 
for the column density variations should lie outside the BLR.     
Note  that such  clouds are quite different from the 
 molecular clouds observed in our Galaxy which 
have a much larger diameter  
(several light years) and a much smaller density (typically 
 $10^{2}~\rm cm^{-3}$).   
 
 One possible problem  with the cloud scenario 
 is the observed anti-correlation between the column density 
 and the unobscured luminosity (Fig.3).  
Indeed, in the simple cloud model there should be no 
 such anti-correlation  
 as the {\it unobscured} luminosity  is only driven by changes 
 in the accretion rate and is hence unrelated to the cloud motion
 and properties. In our case, the 
 increase of the luminosity  while $N_H$ 
 drops could be purely coincidental (eg the luminosity 
 of the source increased while the cloud was moving 
 progressively out of the line of sight).  Indeed the {\it ASCA} point 
 (triangle in Fig 3) suggests that the column density has drifted 
(due to motion of clouds) at least between the {\it ASCA} and 
 {\it RXTE} epochs and 
therefore this correlation should be coincidental.

\section{conclusions}

We model the X-ray spectral variability of Mrk348 
by analyzing 37 RXTE observations 
over a period of 14 months.
The present work  extends our previous study  
(Smith et al. 2001) where 12 observations 
spanning a period of 6 months were  used. 
We find that the column density decreases  by a factor of 3
with increasing flux, confirming our previous results.
These column density variations could arise either due to 
the random drift of clouds within the absorption screen 
or due to photoionisation processes. 

Our modelling shows that photoionization  
alone cannot easily reproduce the observed column density variations. 
In particular, both the {\sl ABSORI} and the {\sl XSTAR} models  
yield a worse fit to the data compared  to the  
double neutral absorber model. 
This is further supported  by the {\it ASCA} observation
which does not show any significant evidence for an ionized edge.
    
 Alternatively  a model which assumes that a  neutral or a weakly 
ionized cloud is moving in front of the source can  
reproduce successfully the column density variations. 
If we assume that the observed $N_H$ variations are caused 
by a single spherical cloud which moves in a circular orbit,
  then, this  cloud should lie in a distance of $>$2 light years from the 
source  having  a density of $>1 \times 10^7 \rm cm^{-3}$. 
 These values are probably characteristic of a region outside the BLR.

Future observations with the XMM mission, which combines large effective area
and excellent spectral resolution will shed more light on the origin
of the obscuring screen in Seyfert 2 galaxies.

\section{acknowledgments}
 We would like to thank the referee G. Risaliti for many useful comments.   
This research has made use of data obtained from the High Energy Astrophysics 
Science Archive Research Center (HEASARC), provided by 
 NASA's Goddard Space Flight Center.

\begin{table*} 
\caption{Best fit parameters}
\begin{tabular}{cccccccc}

\hline 
Model& $\rm N_{H}^1 (cm^{-2})$ & $\rm N_{H}^2 (cm^{-2})$ & covering fraction & $\Gamma$ & $\xi$ & EW(eV) &\x2/dof  \\
\hline 

``standard'' &   $12^{+1}_{-1}-32^{+5}_{-4}$  &-&-& $1.85$ &- & 70-340 & $351/553$   \\

Double absorber (neutral)&  $ 12^{+1}_{-1} $ & $28^{+1}_{-1}  $
 & 0.05-1 & $1.85$ &- &70-300 & $322/551$   \\

Double absorber ($\ sl ABSORI$)&  $ 8^{+1}_{-1} $ &$ 25^{+1}_{-1}  $ 
&- & $1.85 $ & $0-450 $ & 80-410  & $450/551$   \\

Double Absorber (XSTAR)&  $ 7^{+1}_{-1}$ & $27^{+1}_{-1} $ &- & 
$1.85 $ & $0.01-350$ & 75-400  & $460/551$   \\
\hline 

\end{tabular} 

$^{1}$ Neutral column density $\times 10^{22}$ \\
$^{2}$ Ionized or partial covering column density \\

\end{table*}

\end{document}